\documentclass[a4paper,10pt]{article}
\usepackage{amssymb,graphicx}
\begin{document}
\vskip 64pt
\begin{center}
\large{(0,2) Gauged Linear Sigma Model on Supermanifold}
\vspace{16pt}

Yusuke Okame and Mitsuo J. Hayashi\\
\vskip 20pt
Tokai University, 1117 Kitakaname, Hiratsuka, Kanagawa 259-1292, Japan\\
\vskip 16pt
E-mail: mhayashi@keyaki.cc.u-tokai.ac.jp
\end{center}

\thispagestyle{empty}
\vspace{24pt}

\begin{abstract}
 We construct $(0,2)$, $D=2$ gauged linear sigma model on supermanifold with 
both an Abelian and non-Abelian gauge symmetry.
For the purpose of checking the exact supersymmetric (SUSY) invariance of 
the Lagrangian density, 
it is convenient to introduce a new operator $\hat{U}$ for the Abelian gauge 
group. The $\hat{U}$ operator provides consistency conditions for satisfying 
the SUSY invariance. On the other hand, it is not essential to introduce a 
similar operator in order to check the exact SUSY invariance of the 
Lagrangian density of non-Abelian model, contrary to the Abelian one. 
However, we still need a new operator in order to define the (0,2) chirality 
conditions for the (0,2) chiral superfields.
The operator $\hat{U}^{a}$  can be defined from the conditions assuring the
(0,2) supersymmetric invariance of the Lagrangian density in superfield 
formalism for the (0,2) U(N) gauged linear sigma model.

We found consistency conditions for the Abelian gauge group which assure 
(0,2) supersymmetric invariance of Lagrangian density and agree 
with (0,2) chirality conditions for the superpotential. The supermanifold 
$\mathcal{M}^{m|n}$ becomes the super weighted complex projective space 
$WCP^{m-1|n}$ in the U(1) case, which is considered as an example of 
a Calabi-Yau supermanifold. The superpotential $W(\phi,\xi)$ for the 
non-Abelian gauge group satisfies more complex condition for the SU(N) part, 
except the U(1) part of U(N), but does not satisfy a quasi-homogeneous 
condition. This fact implies the need for taking care of constructing 
the Calabi-Yau supermanifold in the SU(N) part. Because more stringent 
restrictions are imposed on the form of the superpotential than in the
U(1) case, the superpotential seems to define a certain kind of new 
supermanifolds which we cannot identify exactly with one of the mathematically 
well defined objects.
\end{abstract}
\newpage
\section{Introduction}
\addtocounter{page}{-1}

Recently, it was reported that the perturbative expansion of the $D=4$, 
$\mathcal{N}=4$ super Yang-Mills theory with the ${\rm U(N)}$ gauge group 
is equivalent to the
instanton expansion of the topological $B$ model for which the target 
space is the Calabi-Yau supermanifold $CP^{3|4}$.
The connection between the topological string theory on supermanifold 
$CP^{3|4}$ and the $D=4$, $\mathcal{N}=4$ super Yang-Mills theory
is established through the explicit calculations of the 
Maximally-Helicity-violating (MHV) amplitude that lead the twistor 
equations \cite{witten1}. Furthermore the methods for calculating many 
types of MHV amplitudes which include loop amplitudes were developed 
\cite{cachazo1} - \cite{brandhuber2}.
%
From these aspects, the Calabi-Yau supermanifold played an important role 
to establish the relation between the super Yang-Mills theory and 
topological B model. Recent works on these relationships 
have investigated of the nature of the Calabi-Yau supermanifold.
The geometry of the Calabi-Yau supermanifold was shown to be related to 
the curvature of the Grassmann even submanifold\cite{rocek1,rocek2} .


The super Landau-Ginzburg mirror symmetry was used to construct the 
correspondence 
between the topological B model on $CP^{3|4}$ as D instanton\cite{witten1}, 
and the topological A model on $CP^{3|3}\times CP^{3|3}$\cite{aganagic1,Ahn1}. 
(These supermanifold are both Calabi-Yau supermanifolds.)
These mirror correspondences were proved by defining the superpotential 
on each case\cite{Kumar1}. 
In fact, the restricted superpotential on A (B) model corresponds to the 
restricted superpotential on B (A) model through the mirror symmetry. 
These restrictions are given by physical symmetries like supersymmetry.
As a simple example, these superpotentials have been shown in the (2,2)  
U(1) gauged linear sigma model,  
because the non-linear sigma model description with Calabi-Yau supermanifold 
is given by the gauged linear sigma model in its infrared limit\cite{Kumar1}. 
Then the restriction of superpotential became equivalent to the (2,2) 
supersymmetric invariance of the total (2,2) Lagrangian density. 

In \cite{seki1}, the Lagrangian density of (2,2) U(1) gauged linear sigma 
model on supermanifold has been constructed. 
The supermanifold then became the Calabi-Yau supermanifold which was 
defined by the Calabi-Yau condition\cite{witten1,Belhaj1}, 
\begin{eqnarray}
\sum_{I}Q_{I}-\sum_{A}q_{A}=0.
\label{q1}
\end{eqnarray}
However, in Ref.\cite{seki1} the (2,2) supersymmetric transformation 
properties of the total (2,2) Lagrangian density seems incomplete, 
because the superpotential term in \cite{seki1} is not exactly closed 
under the (2,2) supersymmetric tansformation when the vector multiplets 
are included. 
If we consider the (2,2) U(1) gauged linear sigma model, the (2,2) 
supersymmetric transformation must include the U(1) vector multiplets in 
its transformation.
Additionally, in Ref.\cite{seki1}, a Lagrangian density of the (0,2) U(1) 
gauged linear sigma model was proposed whose transformation properties 
under the (0,2) supersymmetric transformation was identified by the (0,2) part 
of the (2,2) transformation on supermanifold. However, the U(1) charges 
of each local coordinates must retain the same values. This result means that 
the number of Grassmann even coordinates is equal to the number of Grassmann 
odd coordinates from Eq.(\ref{q1}), as far as we focus on the Calabi-Yau 
supermanifolds which are defined by the mirror symmetric correspondence 
with the super Landau-Ginzburg model. 
In this case, the Calabi-Yau supermanifolds will be ristricted to 
$CP^{m-1|m}$. 

In the present paper, as the first move toward finding out the 
correspondence between the Calabi-Yau supermanifold and the super 
Landau-Ginzburg model, we will concentrate on the construction of a 
consistent theory of the two-dimensional $(0,2)$ U(1) gauged linear 
sigma model on a supermanifold. 
The Lagrangian density of this model becomes (0,2) supersymmetric invariant 
under the corrected (0,2) supersymmetry which includes the vector multiplets.
Then we obtain the restrictions on the superpotential which assure the 
(0,2) supersymmetric invarinace of Lagrangian density.
Furthermore, the conditions define the more general form of the 
Calabi-Yau supermanifold, such as $WCP^{m-1|n}$, by using the newly 
introduced operator $\hat{U}$.
Next, we will construct a consistent (0,2) U(N) gauged linear sigma model 
on supermanifold.
We will show that the restrictions on the superpotential are similar to the 
U(1) gauged linear sigma model for the U(1) part of U(N), while for the 
SU(N) part
the restrictions seem to be stronger than in the U(1) gauged liear sigma 
model on the supermanifolds.

In Section 2, we define supermultiplets of the $D=2$, $(0,2)$ U(1) gauged 
linear sigma model
and construct the Lagrangian densities on a supermanifold, where we 
introduce a new operator in order to distinguish the U(1) charges of local 
coordinates on the supermanifold.
In Section 3, we derive the $(0,2)$ supersymmetric invariance of the 
Lagrangian densities defined in Section 2, and obtain the restriction 
imposed on the superpotential.
We explicitly define the new operator assumed in Section 2 and describe 
the implication of this new operator on the $(0,2)$ supersymmetric 
invariance of the theory. 
In Section 4, we extend the gauge group to the non-Abelian case and 
construct the $D=2$, $(0,2)$ U(N) gauged linear sigma model Lagrangian 
densities on a supermanifold.
In Section 5, the $(0,2)$ supersymmetric invariance is verified on the 
model constructed in Section 4. 
Then we obtain the restriction on the superpotential in the U(N) gauge group.
In Section 6, the operator introduced in Section 3 is extended to the 
non-Abelian U(N) gauge transformation and is shown in relation to 
the $(0,2)$ supersymmetry invariance.

In Section 7, we summarize and discuss our constructions of the 
$D=2$, $(0,2)$ gauged linear sigma models. Our notations are 
the same as those of \cite{wess1}.

\section{$(2,2)$ and $(0,2)$ Supermultiplets}

By introducing several (0,2) superfields, we construct the total (0,2) 
Lagrangian density by a method similar as used in Ref.\cite{seki1}. 
However, in the original method it is impossible to assign 
different values of the 
U(1) charge to each local coordinate. In this section we solve this problem
by introducing a new operator $\hat{U}$, by which it is possible to 
assign different values of the U(1) charge to each local coordinate, 
and construct the more general form of (0,2) Lagrangian density. 
Furthermore, by using the new operator $\hat{U}$, we will obtain the 
more general Calabi-Yau supermanifold, where the number of 
Grassmann even local coordinates and Grassmann odd local coordinates is 
different. This distinction was not made in the method of Ref.\cite{seki1}.

The $D=2$, $\mathcal{N}=2$ superfields are defined on the (2,2) superspace.
We herein redefine these superfields on the (0,2) superspace and 
construct the (0,2) Lagrangian density by using a new operator $\hat{U}$.

In $D=2$, the $(2,2)$ Grassmann even chiral superfield 
$\Phi_{\left(2,2\right)}$ and the $(2,2)$ Grassmann odd chiral 
superfield $\Xi_{\left(2,2\right)}$ are defined as:
\begin{eqnarray}
&&\hspace{-0.5cm}\Phi^{I}_{\left(2,2\right)}
\nonumber\\
&&\hspace{-0.9cm}=\phi^{I}
+\sqrt{\mathstrut 2}\left(\theta^{+}\psi_{+}^{I}
+\theta^{-}\psi_{-}^{I}\right)
+2\theta^{+}\theta^{-}F^{I}
-i\theta^{-}\overline{\theta}^{-}\partial_{-}\phi^{I}
-i\theta^{+}\overline{\theta}^{+}\partial_{+}\phi^{I}
\nonumber\\
&&\hspace{-0.5cm}-\sqrt{\mathstrut 2}i\theta^{+}\theta^{-}\overline{\theta}^{-}\partial_{-}\psi_{+}^{I}
+\sqrt{\mathstrut 2}i\theta^{+}\theta^{-}\overline{\theta}^{+}\partial_{+}\psi_{-}^{I}-\theta^{+}\theta^{-}\overline{\theta}^{-}\overline{\theta}^{+}\partial_{-}\partial_{+}\phi^{I},
\\
&&\hspace{-0.5cm}\Xi^{A}_{\left(2,2\right)}
\nonumber\\
&&\hspace{-0.9cm}=\xi^{A}
+\sqrt{\mathstrut 2}\left(\theta^{+}b_{+}^{A}+\theta^{-}b_{-}^{A}\right)
+2\theta^{+}\theta^{-}\chi^{A}
-i\theta^{-}\overline{\theta}^{-}\partial_{-}\xi^{A}
-i\theta^{+}\overline{\theta}^{+}\partial_{+}\xi^{A}
\nonumber\\
&&\hspace{-0.5cm}-\sqrt{\mathstrut 2}i\theta^{+}\theta^{-}\overline{\theta}^{-}\partial_{-}b_{+}^{A}
+\sqrt{\mathstrut 2}i\theta^{+}\theta^{-}\overline{\theta}^{+}\partial_{+}b_{-}^{A}
-\theta^{+}\theta^{-}\overline{\theta}^{-}\overline{\theta}^{+}\partial_{-}\partial_{+}\xi^{A},
\end{eqnarray}
where $\mu=0,3$, $g_{\mu\nu}={\rm diag}(-1,+1)$, and $\partial_{\pm}=\partial_{0}\pm\partial_{3}$ \cite{seki1}. 
The supermanifold is defined on $\mathcal{M}^{m|n}$, $(I=1,\cdots,m$, $A=1,\cdots,n)$.
For the $(2,2)$ chiral superfield, we introduce the operator $\hat{U}$, which satisfies the following relations:
\begin{eqnarray}
\begin{array}{rcl}
&&\hat{U}\Phi^{I}_{\left(2,2\right)}=Q_{I}\Phi^{I}_{\left(2,2\right)},\hspace{13pt}\hat{U}\overline{\Phi}^{I}_{\left(2,2\right)}=-Q_{I}\overline{\Phi}^{I}_{\left(2,2\right)},\\
&&\hat{U}\Xi^{A}_{\left(2,2\right)}=q_{A}\Xi^{A}_{\left(2,2\right)},\hspace{13pt}\hat{U}\overline{\Xi}^{A}_{\left(2,2\right)}=-q_{A}\overline{\Xi}^{A}_{\left(2,2\right)},
\end{array}
\label{Ucharge1}
\end{eqnarray}
where $Q_{I}$ and $q_{A}$ are the ${\rm U(1)}$ charges of $\Phi^{I}_{\left(2,2\right)}$ and $\Xi^{A}_{\left(2,2\right)}$, respectively, 
and the $\hat{U}$ operator is considered to define the ${\rm U(1)}$ charges of 
the superfields.
We assume that $\hat{U}$ is a Grassmann even operator that satisfies:
\begin{eqnarray}
\begin{array}{rcl}
&&\left[\hat{U},\theta^{\alpha}\right]=\left[\hat{U},\overline{\theta}^{\alpha}\right]=0,\\
&&\left[\hat{U},\frac{\partial}{\partial\theta^{\alpha}}\right]=\left[\hat{U},\frac{\partial}{\partial\overline{\theta}^{\alpha}}\right]=0,
\end{array}
\label{hat{U}1}
\end{eqnarray}
where $\alpha=\pm$.
We define the covariant derivative of the $(0,2)$ supersymmetric transformation by incorporating the $\hat{U}$ and gauge fields $v_{\mu}$,
\begin{eqnarray}
&&{\cal D}_{+}\equiv e^{-\Psi\hat{U}}\left(\frac{\partial}{\partial\theta^{+}}-i\overline{\theta}^{+}\partial_{+}\right)e^{\Psi\hat{U}},
\label{kyouhen1}
\end{eqnarray}
where $v_{\pm}=v_{0}\pm v_{3}$ and $\Psi=\theta^{+}\overline{\theta}^{+}v_{+}$.
The $(0,2)$ super charges are defined by incorporating the $\hat{U}$ operator and the gauge fields $v_{\mu}$ as:
\begin{eqnarray}
&&{\cal Q}_{+}\equiv e^{\Psi\hat{U}}\left(\frac{\partial}{\partial\theta^{+}}+i\overline{\theta}^{+}\partial_{+}\right)e^{-\Psi\hat{U}}.
\label{cyoutaisyou1}
\end{eqnarray}
We now consider the $(0,2)$ case.
The (0,2) chirality conditions are defined by using Eq. (\ref{kyouhen1}) for arbitrary functions $F(x_{\mu},\theta^{+},\overline{\theta}^{+})$ and 
$\overline{F}(x_{\mu},\theta^{+},\overline{\theta}^{+})$ on the (0,2) superspace:
\begin{eqnarray}
{\cal D}_{+}\overline{F}=\overline{\cal D}_{+}F=0.
\label{N=(0,2)chiral condition}
\end{eqnarray}

We can define the $(0,2)$ chiral superfields that satisfy Eq. (\ref{N=(0,2)chiral condition}) from the $(2,2)$ chiral multiplets 
by imposing restrictions $\theta^{-}=\overline{\theta}^{-}=0$ \cite{seki1}.
\begin{eqnarray}
&&\hspace{-0.7cm}\Phi^{I}_{\left(0,2\right)}\equiv\Phi^{I}_{\left(2,2\right)}e^{Q_{I}\Psi}\biggl|_{\theta^{-}=\overline{\theta}^{-}=0},
\label{Phi2}\\
&&\hspace{-0.7cm}\Xi^{A}_{\left(0,2\right)}\equiv\Xi^{A}_{\left(2,2\right)}e^{q_{A}\Psi}\biggl|_{\theta^{-}=\overline{\theta}^{-}=0},
\label{Xi2}
\end{eqnarray}
where the covariant derivatives for the ${\rm U(1)}$ gauge transformation are given by
\begin{eqnarray}
&&D_{\mu}=\partial_{\mu}+iv_{\mu}\hat{U}.
\end{eqnarray}

Since Eqs. (\ref{Phi2}) and (\ref{Xi2}) satisfy the $(0,2)$ chirality conditions, they are the $(0,2)$ chiral superfields:
\begin{eqnarray}
&&\overline{\cal D}_{+}\Phi^{I}_{\left(0,2\right)}={\cal D}_{+}\overline{\Phi}^{I}_{\left(0,2\right)}=0,\\
&&\overline{\cal D}_{+}\Xi^{A}_{\left(0,2\right)}={\cal D}_{+}\overline{\Xi}^{A}_{\left(0,2\right)}=0.
\end{eqnarray}

Next, we define the Lagrangian density ${\cal L}_{kin.}$ for the $(0,2)$ chiral superfields.
Denoting the $\theta^{-}\overline{\theta}^{-}$ term of the $(2,2)$ vector 
superfields as ${\cal V}$, we have
\begin{eqnarray}
{\cal V}=iv_{-}+2\theta^{+}\overline{\lambda}_{-}+2\overline{\theta}^{+}\lambda_{-}+2i\theta^{+}\overline{\theta}^{+}D.
\label{cal V1}
\end{eqnarray}
The ${\rm U(1)}$ charge for ${\cal V}$ is assumed to be zero, i.e.,
\begin{eqnarray}
\hat{U}{\cal V}=0.
\label{Ucharge'1}
\end{eqnarray}

From the assumptions on ${\rm U(1)}$ charges for $(0,2)$ chiral superfields in Eqs. (\ref{Ucharge1}) and (\ref{Ucharge'1}), we obtain  
\begin{eqnarray}
\begin{array}{rcl}
&&\hat{U}\Phi^{I}_{\left(0,2\right)}=Q_{I}\Phi^{I}_{\left(0,2\right)},\hspace{13pt}\hat{U}\overline{\Phi}^{I}_{\left(0,2\right)}=-Q_{I}\overline{\Phi}^{I}_{\left(0,2\right)},
\\
&&\hat{U}\Xi^{A}_{\left(0,2\right)}=q_{A}\Xi^{A}_{\left(0,2\right)},\hspace{13pt}\hat{U}\overline{\Xi}^{A}_{\left(0,2\right)}=-q_{A}\overline{\Xi}^{A}_{\left(0,2\right)}.
\end{array}
\end{eqnarray}

Using $\mathcal{V}$, we define the covariant derivative for gauge transformation:
\begin{eqnarray}
{\cal D}_{0}-{\cal D}_{3}\equiv\partial_{-}+{\cal V}\hat{U}.
\label{cal D_{0}-D_{3}}
\end{eqnarray}

From the (0,2) chiral superfields and Eq. (\ref{cal D_{0}-D_{3}}), ${\cal L}_{kin.}$ is given by
\begin{eqnarray}
&&{\cal L}_{kin.}=\frac{i}{2}\int d\theta^{+}d\overline{\theta}^{+}\Bigg[\sum_{I}\overline{\Phi}^{I}_{\left(0,2\right)}\left({\cal D}_{0}-{\cal D}_{3}\right)\Phi^{I}_{\left(0,2\right)}
\nonumber\\
&&\hspace{3.6cm}+\sum_{A}\overline{\Xi}^{A}_{\left(0,2\right)}\left({\cal D}_{0}-{\cal D}_{3}\right)\Xi^{A}_{\left(0,2\right)}\Bigg].
\label{{cal L}_kin.}
\end{eqnarray}
Next, we will define the Lagrangian density ${\cal L}_{gauge}$ and
the Fayet-Iliopoulos 
(FI) term ${\cal L}_{D,\theta}$ for the vector superfield $\mathcal{V}$. 
The gauge invariant field strength $\Upsilon$ is defined as
\begin{eqnarray}
&&\Upsilon\equiv\overline{\cal D}_{+}{\cal V}+\theta^{+}\partial_{-}v_{+}.
\label{Upsilon1}
\end{eqnarray}

From Eq. (\ref{Upsilon1}), the kinetic Lagrangian density ${\cal L}_{gauge}$ of this gauge multiplet is given as
\begin{eqnarray}
&&\hspace{-0.5cm}{\cal L}_{gauge}=\frac{1}{8e^2}\int d\theta^{+}d\overline{\theta}^{+}\overline{\Upsilon}\Upsilon,
\label{{cal L}_gauge}
\end{eqnarray}
and the FI term is
\begin{eqnarray}
&&{\cal L}_{D,\theta}=\frac{t}{4}\int d\theta^{+}\Upsilon\hspace{0.1cm}\Biggl|_{\overline{\theta}^{+}=0}+\frac{\overline{t}}{4}\int d\overline{\theta}^{+}\overline{\Upsilon}\hspace{0.1cm}\Biggl|_{\theta^{+}=0},
\label{{cal L}_D,theta}
\end{eqnarray}
with the FI parameter $t=ir+\theta/(2\pi)$.

Since the field components in Eq. (\ref{Upsilon1}) contain part of the 
(2,2) vector supermultiplet,
the residual field components should be introduced into the theory by 

\begin{eqnarray}
&&\Omega\equiv\tau+\sqrt{\mathstrut 2}i\theta^{+}\overline{\omega}_{+}-i\theta^{+}\overline{\theta}^{+}\partial_{+}\tau,
\label{Omega1}
\end{eqnarray}
where Eq. (\ref{Omega1}) is a Grassmann even superfield and assumed to be 
chargeless, i.e.,
\begin{eqnarray}
\hat{U}\Omega=\hat{U}\overline{\Omega}=0\label{Ucharge2}.
\end{eqnarray}

From Eq. (\ref{Omega1}), we can define the Lagrangian density ${\cal L}_{\Omega}$ as
\begin{eqnarray}
&&\hspace{-1.6cm}{\cal L}_{\Omega}=\frac{i}{2e^{2}}\int d\theta^{+}d\overline{\theta}^{+}\overline{\Omega}\partial_{-}\Omega.
\label{{cal L}_Omega}
\end{eqnarray}

In order to construct the $(0,2)$ superpotential consitently, we introduce some $(0,2)$ chiral superfield valued functions 
$E_{a}(\Phi_{\left(0,2\right)},\Omega)$,
$\tilde{E}_{\tilde{a}}(\Xi_{\left(0,2\right)},\Omega)$
where the indices $a$ and $\tilde{a}$ denote Grassmann even and Grassmann odd, respectively.
In addition, the other $(0,2)$ superfields are introduced as 
\begin{eqnarray}
&&\Lambda_{-a}^{'}\equiv\lambda_{-a}-\sqrt{\mathstrut 2}\theta^{+}G_{a}-i\theta^{+}\overline{\theta}^{+}\partial_{+}\lambda_{-a},
\label{Lambda_a'1}
\\
&&\tilde{\Lambda}_{-\tilde{a}}^{'}\equiv\tilde{\lambda}_{-\tilde{a}}-\sqrt{\mathstrut 2}\theta^{+}\tilde{G}_{\tilde{a}}-i\theta^{+}\overline{\theta}^{+}\partial_{+}\tilde{\lambda}_{-\tilde{a}}.
\label{tilde{Lambda}_a'1}
\end{eqnarray}
The ${\rm U(1)}$ charges for the fields in Eqs. (\ref{Lambda_a'1}) and (\ref{tilde{Lambda}_a'1}) are assumed as
\begin{eqnarray}
\begin{array}{rcl}
&&\hat{U}\Lambda_{-a}^{'}=\alpha_{a}\Lambda_{-a}^{'},
\hspace{13pt}
\hat{U}\overline{\Lambda}_{-a}^{'}=-\alpha_{a}\overline{\Lambda}_{-a}^{'},
\\
&&\hat{U}\tilde{\Lambda}_{-\tilde{a}}^{'}=\beta_{\tilde{a}}\tilde{\Lambda}_{-\tilde{a}}^{'},
\hspace{13pt}
\hat{U}\overline{\tilde{\Lambda}}_{-\tilde{a}}^{'}=-\beta_{\tilde{a}}\overline{\tilde{\Lambda}}_{-\tilde{a}}^{'}.
\end{array}
\label{Ucharge3}
\end{eqnarray}

Moreover, by Eqs. (\ref{Lambda_a'1}) and (\ref{tilde{Lambda}_a'1}), 
we define new fields as
\begin{eqnarray}
&&\Lambda_{-a}\equiv\Lambda_{-a}^{'}-\sqrt{\mathstrut 2}\overline{\theta}^{+}E_{a}(\Phi,\Omega),
\label{Lambda_a1}
\\
&&\tilde{\Lambda}_{-\tilde{a}}\equiv\tilde{\Lambda}_{-\tilde{a}}^{'}-\sqrt{\mathstrut 2}\overline{\theta}^{+}\tilde{E}_{\tilde{a}}(\Xi,\Omega).
\label{tilde{Lambda}_a1}
\end{eqnarray}
We can then define the $(0,2)$ superfields as:
\begin{eqnarray}
&&\hspace{-1cm}\Lambda_{-a\left(0,2\right)}\equiv\Lambda_{-a}e^{\alpha_{a}\Psi},
\label{Lambda_a{02}1}
\\
&&\hspace{-1cm}\tilde{\Lambda}_{-\tilde{a}\left(0,2\right)}\equiv\tilde{\Lambda}_{-\tilde{a}}e^{\beta_{\tilde{a}}\Psi},
\label{tilde{Lambda}_a{02}1}
\end{eqnarray}
by using Eqs. (\ref{Ucharge3}), (\ref{Lambda_a1}) and (\ref{tilde{Lambda}_a1}).
We then obtain the Lagrangian density ${\cal L}_{\Lambda}$ from Eqs. (\ref{Lambda_a{02}1}) and (\ref{tilde{Lambda}_a{02}1}).
\begin{eqnarray}
&&\hspace{-1cm}{\cal L}_{\Lambda}
=\frac{1}{2}\int d\theta^{+}d\overline{\theta}^{+}\Biggl[\sum_{a}\overline{\Lambda}_{-a\left(0,2\right)}\Lambda_{-a\left(0,2\right)}
+\sum_{\tilde{a}}\overline{\tilde{\Lambda}}_{-\tilde{a}\left(0,2\right)}\tilde{\Lambda}_{-\tilde{a}\left(0,2\right)}\Biggr].
\label{{cal L}_Lambda}
\end{eqnarray}

We need more $(0,2)$ chiral superfield valued functions 
$J^{a}(\Phi_{\left(0,2\right)},\Xi_{\left(0,2\right)})$,
$\tilde{J}^{\tilde{a}}(\Phi_{\left(0,2\right)},\Xi_{\left(0,2\right)})$.
The ${\rm U(1)}$ charges for these fields are assumed as
\begin{eqnarray}
&&\hat{U}J^{a}(\phi,\xi)=\sum_{I}\hat{U}\phi^{I}\frac{\partial J^{a}(\phi,\xi)}{\partial\phi^{I}}
+\sum_{A}\hat{U}\xi^{A}\frac{\partial J^{a}(\phi,\xi)}{\partial\xi^{A}},
\label{Ucharge7}
\\
&&\hat{U}\tilde{J}^{\tilde{a}}(\phi,\xi)=\sum_{I}\hat{U}\phi^{I}\frac{\partial\tilde{J}^{\tilde{a}}(\phi,\xi)}{\partial\phi^{I}}
+\sum_{A}\hat{U}\xi^{A}\frac{\partial\tilde{J}^{\tilde{a}}(\phi,\xi)}{\partial\xi^{A}}.
\label{Ucharge9}
\end{eqnarray}
Here we impose the following restrictions on the fields $E_{a}(\phi,\tau)$, $\tilde{E}_{\tilde{a}}(\phi,\tau)$ and $J^{a}(\phi,\xi)$, $\tilde{J}^{\tilde{a}}(\phi,\xi)$:
\begin{eqnarray}
&&\sum_{a}E_{a}(\phi,\tau)J^{a}(\phi,\xi)
+\sum_{\tilde{a}}\tilde{E}_{\tilde{a}}(\xi,\tau)\tilde{J}^{\tilde{a}}(\phi,\xi)=0.
\label{E_aJ^a1}
\end{eqnarray}
From these restrictions, we can obtain 
the (0,2) chirality conditions
\begin{eqnarray}
&&\overline{\cal D}_{+}\left(\sum_{a}\Lambda_{-a\left(0,2\right)}J^{a}(\Phi_{\left(0,2\right)},\Xi_{\left(0,2\right)})
+\sum_{\tilde{a}}\tilde{\Lambda}_{-\tilde{a}\left(0,2\right)}\tilde{J}^{\tilde{a}}(\Phi_{\left(0,2\right)},\Xi_{\left(0,2\right)})\right)
\nonumber\\
&&\hspace{-0.4cm}={\cal D}_{+}\Biggl(\sum_{a}\overline{J}^{a}(\overline{\Phi}_{\left(0,2\right)},\overline{\Xi}_{\left(0,2\right)})\overline{\Lambda}_{-a\left(0,2\right)}
+\sum_{\tilde{a}}\overline{\tilde{J}}^{\tilde{a}}(\overline{\Phi}_{\left(0,2\right)},\overline{\Xi}_{\left(0,2\right)})\overline{\tilde{\Lambda}}_{-\tilde{a}\left(0,2\right)}\Biggr)
\nonumber\\
&&\hspace{-0.4cm}=0,
\label{overline{JLambda}1}
\end{eqnarray}
which define $(0,2)$ chiral superfields and provide the Lagrangian density ${\cal L}_{J}$ as follows:
\begin{eqnarray}
&&\hspace{-0.8cm}{\cal L}_{J}=\frac{1}{\sqrt{\mathstrut 2}}\int d\theta^{+}\Biggl[\sum_{a}\Lambda_{-a\left(0,2\right)}J^{a}(\Phi_{\left(0,2\right)},\Xi_{\left(0,2\right)})
\nonumber\\
&&\hspace{2.2cm}
+\sum_{\tilde{a}}\tilde{\Lambda}_{-\tilde{a}\left(0,2\right)}\tilde{J}^{\tilde{a}}(\Phi_{\left(0,2\right)},\Xi_{\left(0,2\right)})\Biggr]\Biggl|_{\overline{\theta}^{+}=0}
\nonumber\\
&&\hspace{0.3cm}+\frac{1}{\sqrt{\mathstrut 2}}\int d\overline{\theta}^{+}\Biggl[\sum_{a}\overline{J}^{a}(\overline{\Phi}_{\left(0,2\right)},\overline{\Xi}_{\left(0,2\right)})\overline{\Lambda}_{-a\left(0,2\right)}
\nonumber\\
&&\hspace{2.5cm}+\sum_{\tilde{a}}\overline{\tilde{J}}^{\tilde{a}}(\overline{\Phi}_{\left(0,2\right)},\overline{\Xi}_{\left(0,2\right)})\overline{\tilde{\Lambda}}_{-\tilde{a}\left(0,2\right)}\Biggr]\Biggr|_{\theta^{+}=0}.
\label{{cal L}_J}
\end{eqnarray}

We now describe the correspondences between the $(2,2)$ field components 
and the $(0,2)$ field components.
First, the following differential operator is defined:
\begin{eqnarray}
&&{\cal D}_{-}\equiv e^{-\Pi\hat{U}}\left(\frac{\partial}{\partial\theta^{-}}-i\overline{\theta}^{-}\partial_{-}\right)e^{\Pi\hat{U}},
\label{kyouhen3}
\end{eqnarray}
where $\Pi=\theta^{-}\overline{\theta}^{-}v_{-}$.
We can then find the following relations:
\begin{eqnarray}
&&\Lambda_{-a}^{'}e^{\alpha_{a}\Psi}
=\frac{1}{\sqrt{\mathstrut 2}}{\cal D}_{-}\left(\Phi_{\left(2,2\right)}^{I}e^{Q_{I}\Psi}\right)\biggl|_{\theta^{-}=\overline{\theta}^{-}=0},
\label{Lambdae2}
\\
&&\tilde{\Lambda}_{-\tilde{a}}^{'}e^{\beta_{\tilde{a}}\Psi}
=\frac{1}{\sqrt{\mathstrut 2}}{\cal D}_{-}\left(\Xi_{\left(2,2\right)}^{A}e^{q_{A}\Psi}\right)\biggl|_{\theta^{-}=\overline{\theta}^{-}=0}.
\label{tilde{Lambda}e2}
\end{eqnarray}
From these relations, it is shown that $I=a$, $A=\tilde{a}$ for indices and $Q_{I}=\alpha_{a}$, $q_{A}=\beta_{\tilde{a}}$ for ${\rm U(1)}$ charges.
The exact correspondences between the field components of the $(2,2)$ chiral superfield and those of the $(0,2)$ superfields are given as follows:
\begin{eqnarray}
\begin{array}{rcl}
&&\lambda_{-a}=\psi_{-}^{I},\hspace{13pt}
\tilde{\lambda}_{-\tilde{a}}=b_{-}^{A},
\\
&&G_{a}=F^{I},\hspace{13pt}
\tilde{G}_{\tilde{a}}=\chi^{A}.
\end{array}
\label{22 021}
\end{eqnarray}

In order to find the corresponding relations between the $(2,2)$ superfields and the $(0,2)$ superfields, the products of the $(2,2)$ chiral superfields and the $(2,2)$ twist chiral superfields are shown.
The $(2,2)$ twist chiral superfield is defined as: 
\begin{eqnarray}
&&\hspace{-1cm}\Sigma=\sigma
+\sqrt{\mathstrut 2}i\theta^{+}\overline{\lambda}_{+}
-\sqrt{\mathstrut 2}i\overline{\theta}^{-}\lambda_{-}
+\sqrt{\mathstrut 2}\theta^{+}\overline{\theta}^{-}\left(D-iv_{03}\right)
+i\theta^{-}\overline{\theta}^{-}\partial_{-}\sigma
\nonumber\\
&&\hspace{-0.3cm}-i\theta^{+}\overline{\theta}^{+}\partial_{+}\sigma
-\sqrt{\mathstrut 2}\theta^{+}\theta^{-}\overline{\theta}^{-}\partial_{-}\overline{\lambda}_{+}
+\sqrt{\mathstrut 2}\theta^{+}\overline{\theta}^{-}\overline{\theta}^{+}\partial_{+}\lambda_{-}
\nonumber\\
&&\hspace{-0.3cm}+\theta^{+}\theta^{-}\overline{\theta}^{-}\overline{\theta}^{+}\partial_{-}\partial_{+}\sigma.
\end{eqnarray}

We find the correspondences as follows:
\begin{eqnarray}
&&\overline{\cal D}_{+}\Lambda_{-a\left(0,2\right)}
=2Q_{I}\Sigma\Phi^{I}_{\left(2,2\right)}e^{Q_{I}\Psi}\biggl|_{\theta^{-}=\overline{\theta}^{-}=0},
\label{DLambda2}
\\
&&\overline{\cal D}_{+}\tilde{\Lambda}_{-\tilde{a}\left(0,2\right)}
=2q_{A}\Sigma\Xi^{A}_{\left(2,2\right)}e^{q_{A}\Psi}\biggl|_{\theta^{-}=\overline{\theta}^{-}=0},
\label{Dtilde{Lambda}2}
\end{eqnarray}
where we assumed the following relations:
\begin{eqnarray}
&&\hat{U}E_{a}(\phi,\tau)
=\sum_{I}\hat{U}\phi^{I}\frac{\partial E_{a}(\phi,\tau)}{\partial\phi^{I}},
\label{Ucharge'2}
\\
&&\hat{U}\tilde{E}_{\tilde{a}}(\xi,\tau)
=\sum_{A}\hat{U}\xi^{A}\frac{\partial\tilde{E}_{\tilde{a}}(\xi,\tau)}{\partial\xi^{A}}.
\label{Ucharge'4}
\end{eqnarray}
The correspondences between the field components of the $(2,2)$ chiral 
superfields and the $(0,2)$ superfields are derived as 
\begin{eqnarray}
\begin{array}{rcl}
&&E_{a}(\phi,\tau)
=\sqrt{\mathstrut 2}Q_{I}\sigma\phi^{I},
\\
&&\tilde{E}_{\tilde{a}}(\xi,\tau)
=\sqrt{\mathstrut 2}q_{A}\sigma\xi^{A},
\\
&&\tau=\sigma,\hspace{13pt}
\omega_{+}=\lambda_{+}.
\end{array}
\label{22 022}
\end{eqnarray}


Finally, we present the following relations between the field components 
of the $(0,2)$ superfields
$J^{a}(\Phi_{\left(0,2\right)},\Xi_{(0,2)})$, 
$\tilde{J}^{\tilde{a}}(\Phi_{\left(0,2\right)},\Xi_{\left(0,2\right)})$ and those of the $(2,2)$ superfields by using the superpotential $W$ as
\begin{eqnarray}
\begin{array}{rcl}
&&J^{a}(\phi,\xi)
=\frac{\partial W(\phi,\xi)}{\partial\phi^{I}},
\\
&&\tilde{J}^{\tilde{a}}(\phi,\xi)
=\frac{\partial W(\phi,\xi)}{\partial\xi^{A}}.
\end{array}
\label{22 023}
\end{eqnarray}

We have shown that the total $(0,2)$ Lagrangian density ${\cal L}_{\left(0,2\right)}$ is obtained from Eqs. (\ref{{cal L}_kin.}), (\ref{{cal L}_gauge}), (\ref{{cal L}_D,theta}), (\ref{{cal L}_Omega}), (\ref{{cal L}_Lambda}), and (\ref{{cal L}_J}) as follows:
\begin{eqnarray}
{\cal L}_{\left(0,2\right)}={\cal L}_{kin.}+{\cal L}_{gauge}+{\cal L}_{D,\theta}+{\cal L}_{\Omega}+{\cal L}_{\Lambda}+{\cal L}_{J}.
\label{{cal L}_02}
\end{eqnarray}
As a result, by the method of using the operator $\hat{U}$, the $(0,2)$ action of the total Lagrangin density of Eq. (\ref{{cal L}_02}) agrees exactly with the $(2,2)$ action $S_{\left(2,2\right)}$ in \cite{seki1}, because of the correspondences in Eqs. (\ref{22 021}), (\ref{22 022}), and (\ref{22 023}).
By using the new operator $\hat{U}$, unlike in the method of Ref.\cite{seki1} 
where it is impossible to have different values of U(1) charges, we could 
assign different values of U(1) charges to each local coordinate, and 
provide a more general (0,2) Lagrangian density. These result will lead to a
more general Calabi-Yau supermanifold which has a different number of 
even local coordinates and odd local coordinates, as will be shown in 
later sections.

\section{$(0,2)$ Supersymmetric Transformations and Invariance of 
Lagrangian Densities}

In this section we will show the $(0,2)$ supersymmetric transformation 
properties of the field components, and prove the $(0,2)$ supersymmetric invariances of the Lagrangian densities introduced in section 2, up to the total derivatives.

The (2,2) supersymmetric transformation property of the total (2,2) Lagrangian density has been indicated in Ref.\cite{seki1}.
However the (0,2) supersymmetric transformation property of the total (0,2) Lagrangian density, in which each local coordinate has the same U(1) charge, has not yet been explicitly indicated.
We are able to find the (0,2) supersymmetric transformation property of 
the (0,2) Lagrangian density by looking at the (2,2) supersymmetric 
transformation property of the (2,2) Lagrangian density indirectly.
This is different from the method of Ref.\cite{seki1}, where the (0,2) 
supersymmetric transformation property of the superpotential term is 
calculated indirectly, and the supersymmetric transformation of U(1) 
vector multiplets and the U(1) gauge transformation is not included. 
Therefore, by assigning different values of U(1) charges to each 
local coordinate and assigning the correct (0,2) supersymmetric transformation 
property to the superpotential term, we define the (0,2) supersymmetric transformation operator by using the new operator $\hat{U}$, and we will verify the (0,2) supersymmetric transformation property of the total (0,2) 
Lagrangian density.

First, we define an operation of the $(0,2)$ supersymmetric transformations
from Eq. (\ref{cyoutaisyou1}):
\begin{eqnarray}
\delta_{\left(0,2\right)}=-\epsilon_{-}{\cal Q}_{+}+\overline{\epsilon}_{-}\overline{\cal Q}_{+}.
\label{cyoutaisyou3}
\end{eqnarray}

We can then derive the $(0,2)$ supersymmetric transformation properties of the
field components,
and can show that they match the $(2,2)$ supersymmetric transformation 
properties by using the correspondence relations 
of Eqs. (\ref{22 021}), (\ref{22 022}), and (\ref{22 023}).
The $(0,2)$ supersymmetric transformation properties for the field components 
of the $(2,2)$ Grassmann even chiral superfield are given by
\begin{eqnarray}
\begin{array}{rcl}
&&\delta_{\left(0,2\right)}\phi^{I}
=-{\sqrt{\mathstrut 2}}\epsilon_{-}\psi_{+}^{I},
\\
&&\delta_{\left(0,2\right)}\psi_{+}^{I}
={\sqrt{\mathstrut 2}}i\overline{\epsilon}_{-}D_{+}\phi^{I},
\\
&&\delta_{\left(0,2\right)}\psi_{-}^{I}
={\sqrt{\mathstrut 2}}\left(\epsilon_{-}F^{I}
+{\sqrt{\mathstrut 2}}\overline{\epsilon}_{-}Q_{I}\sigma\phi^{I}\right),
\\
&&\delta_{\left(0,2\right)}F^{I}
=-{\sqrt{\mathstrut 2}}\overline{\epsilon}_{-}\left(iD_{+}\psi_{-}^{I}-{\sqrt{\mathstrut 2}}iQ_{I}\overline{\lambda}_{+}\phi^{I}
-{\sqrt{\mathstrut 2}}Q_{I}\psi_{+}^{I}\sigma\right).
\end{array}
\end{eqnarray}
In addition, the $(0,2)$ supersymmetric transformation properties for the field
components of the $(2,2)$ Grassmann odd chiral superfield are:

\begin{eqnarray}
\begin{array}{rcl}
&&\delta_{\left(0,2\right)}\xi^{A}
=-{\sqrt{\mathstrut 2}}\epsilon_{-}b_{+}^{A},
\\
&&\delta_{\left(0,2\right)}b_{+}^{A}
={\sqrt{\mathstrut 2}}i\overline{\epsilon}_{-}D_{+}\xi^{A},
\\
&&\delta_{\left(0,2\right)}b_{-}^{A}
={\sqrt{\mathstrut 2}}\left(\epsilon_{-}\chi^{A}+{\sqrt{\mathstrut 2}}\overline{\epsilon}_{-}q_{A}\sigma\xi^{A}\right),
\\
&&\delta_{\left(0,2\right)}\chi^{A}
=-{\sqrt{\mathstrut 2}}\overline{\epsilon}_{-}\left(iD_{+}b_{-}^{A}
-{\sqrt{\mathstrut 2}}iq_{A}\overline{\lambda}_{+}\xi^{A}
-{\sqrt{\mathstrut 2}}q_{A}b_{+}^{A}\sigma\right).
\end{array}
\end{eqnarray}
The $(0,2)$ supersymmetric transformation properties for the field
components of the $(2,2)$ vector superfield
are given as follows:
\begin{eqnarray}
\begin{array}{rcl}
&&\delta_{\left(0,2\right)}v_{+}
=0,
\\
&&\delta_{\left(0,2\right)}v_{-}
=2i\left(\epsilon_{-}\overline{\lambda}_{-}+\overline{\epsilon}_{-}\lambda_{-}\right),
\\
&&\delta_{\left(0,2\right)}\sigma
=-{\sqrt{\mathstrut 2}}i\epsilon_{-}\overline{\lambda}_{+}
,\\
&&\delta_{\left(0,2\right)}\lambda_{-}
=i\epsilon_{-}\left(D-iv_{03}\right)
,\\
&&\delta_{\left(0,2\right)}\lambda_{+}
={\sqrt{\mathstrut 2}}\epsilon_{-}\partial_{+}\overline{\sigma}
,\\
&&\delta_{\left(0,2\right)}D
=\epsilon_{-}\partial_{+}\overline{\lambda}_{-}
-\overline{\epsilon}_{-}\partial_{+}\lambda_{-}
,
\end{array}
\end{eqnarray}
Where $v_{03}=\partial_{0}v_{3}-\partial_{3}v_{0}$.

The $(0,2)$ supersymmetric transformation properties of the Lagrangian densities $\mathcal{L}_{\left(0,2\right)}$
are derived and expressed by using the $(2,2)$ field components from 
Eqs. (\ref{22 021}), (\ref{22 022}), and (\ref{22 023}).
The actions for $\mathcal{L}_{\left(0,2\right)}$ are supersymmetric 
invariants up to total derivatives.
However, for ${\cal L}_{J}$ in Eq. (\ref{{cal L}_J}), we obtain:
\begin{eqnarray}
&&\delta_{\left(0,2\right)}{\cal L}_{J}
\nonumber\\
&&\hspace{-0.4cm}
=\sqrt{\mathstrut 2}i\overline{\epsilon}_{-}\partial_{+}\Biggl[\sum_{I}\psi_{-}^{I}\frac{\partial W(\phi,\xi)}{\partial\phi^{I}}
+\sum_{A}b_{-}^{A}\frac{\partial W(\phi,\xi)}{\partial\xi^{A}}\Biggr]
\nonumber\\
&&-\sqrt{\mathstrut 2}\overline{\epsilon}_{-}\Biggl[v_{+}\hat{U}\left(\sum_{I}\psi_{-}^{I}\frac{\partial W(\phi,\xi)}{\partial\phi^{I}}
+\sum_{A}b_{-}^{A}\frac{\partial W(\phi,\xi)}{\partial\xi^{A}}\right)
\nonumber\\
&&+\Biggl(\sum_{J}\psi_{+}^{J}\frac{\partial}{\partial\phi^{J}}
+\sum_{B}b_{+}^{B}\frac{\partial}{\partial\xi^{B}}
+i\overline{\lambda}_{+}\frac{\partial}{\partial\sigma}\Biggr)
\Biggl(\sum_{I}\sqrt{\mathstrut 2}Q_{I}\sigma\phi^{I}\frac{\partial W(\phi,\xi)}{\partial\phi^{I}}
\nonumber\\
&&+\sum_{A}\sqrt{\mathstrut 2}q_{A}\sigma\xi^{A}\frac{\partial W(\phi,\xi)}{\partial\xi^{A}}\Biggr)\Biggr]
+(h.c.).
\label{delta{cal L}_J}
\end{eqnarray}
The results of Eq. (\ref{delta{cal L}_J}) imply that the action for $\mathcal{L}_{J}$ is not a supersymmetric invariant,
because the variations consist of non-total derivative terms under the $(0,2)$ supersymmetric transformation.
 
Therefore, we must impose consistency conditions that will assure that ${\cal L}_{J}$ is $(0,2)$ supersymmetric invariant up to total derivatives:
\begin{eqnarray}
&&\sum_{I}Q_{I}\phi^{I}\frac{\partial W(\phi,\xi)}{\partial\phi^{I}}
+\sum_{A}q_{A}\xi^{A}\frac{\partial W(\phi,\xi)}{\partial\xi^{A}}
=0.
\label{zyouken1}
\end{eqnarray}
Equation (\ref{zyouken1}) is the same condition as Eq. (\ref{E_aJ^a1}).
We undestand that Eq. (\ref{E_aJ^a1}) does not only define $(0,2)$ chiral superfields, but also gives a consistency condition that ensures the supersymmetric variation of ${\cal L}_{J}$ to be invariant under $(0,2)$ supersymmetric 
transformations up to total derivatives. 

These restrictions on the superpotential are confirmed by using the 
corrected (0,2) supersymmetric transformation which includes the U(1) 
vector multiplets.
Using the method of Ref.\cite{seki1}, one could not confirm the necessity 
of the restrictions clearly.
However, we in our present method we could indicate explicitly the 
necessity of the restrictions.
It has been reported (Ref.\cite{seki1}) that Eq. (\ref{zyouken1}) is equivalent to the quasi-homogeneous condition $W(\phi^{I},\xi^{A})=W(\lambda^{Q_{I}}\phi^{I},\lambda^{q_{A}}\xi^{A})$ for the superpotential.
Thus, we can use the identification:
\begin{eqnarray}
&&(\phi^{1},\phi^{2},\cdots,\phi^{m}|\xi^{1},\xi^{2},\cdots,\xi^{n})
\nonumber\\
&& \hspace{1cm}\sim (\lambda^{Q_{1}}\phi^{1},\lambda^{Q_{2}}\phi^{2},\cdots,\lambda^{Q_{m}}\phi^{m}|\lambda^{q_{1}}\xi^{1},\lambda^{q_{2}}\xi^{2},\cdots,\lambda^{q_{n}}\xi^{n}),
\end{eqnarray}
where $\lambda\in C^{\times}$.
Namely, the supermanifold $\mathcal{M}^{m|n}$ becomes the super weighted complex projective space $WCP^{m-1|n}$, which can be reproduced using $\hat{U}$.
If we focus on the Calabi-Yau supermanifold corresponding to the super 
Landau-Ginzburg model, we can construct a Calabi-Yau supermanifold, which
is more general in Ref.\cite{seki1} and which has different numbers of 
even coordinates and odd coordinates satisfying Eq. (\ref{q1}). 
%

The formula of the $\hat{U}$ charge operator satisfies the assumptions of Eqs. (\ref{Ucharge1}), (\ref{hat{U}1}), (\ref{Ucharge'1}), (\ref{Ucharge2}), (\ref{Ucharge3}), 
(\ref{Ucharge7}) and (\ref{Ucharge'2}).
The operator $\hat{U}$ is written as follows:
\begin{eqnarray}
&&\hspace{-0.7cm}
\hat{U}=\sum_{I}Q_{I}\Biggl[\phi^{I}\frac{\partial}{\partial\phi^{I}}
+\sum_{\mu}\partial_{\mu}\phi^{I}\frac{\partial}{\partial\left(\partial_{\mu}\phi^{I}\right)}
+\sum_{\mu,\nu}\partial_{\mu}\partial^{\mu}\phi^{I}\frac{\partial}{\partial\left(\partial_{\nu}\partial^{\nu}\phi^{I}\right)}
+\sum_{\alpha=\pm}\Bigg\{\psi^{I}_{\alpha}\frac{\partial}{\partial\psi^{I}_{\alpha}}
\nonumber\\
&&
+\sum_{\mu}\partial_{\mu}\psi^{I}_{\alpha}\frac{\partial}{\partial\left(\partial_{\mu}\psi^{I}_{\alpha}\right)}\Bigg\}
+F^{I}\frac{\partial}{\partial F^{I}}
\nonumber\\
&&+\sum_{A}q_{A}\Biggl[\xi^{A}\frac{\partial}{\partial\xi^{A}}
+\sum_{\mu}\partial_{\mu}\xi^{A}\frac{\partial}{\partial\left(\partial_{\mu}\xi^{A}\right)}
+\sum_{\mu,\nu}\partial_{\mu}\partial^{\mu}\xi^{A}\frac{\partial}{\partial\left(\partial_{\nu}\partial^{\nu}\xi^{A}\right)}
+\sum_{\alpha=\pm}\Bigg\{b_{\alpha}^{A}\frac{\partial}{\partial b_{\alpha}^{A}}
\nonumber\\
&&
+\sum_{\mu}\partial_{\mu}b_{\alpha}^{A}\frac{\partial}{\partial\left(\partial_{\mu}b_{\alpha}^{A}\right)}\Bigg\}
+\chi^{A}\frac{\partial}{\partial\chi^{A}}
+(h.c.).
\label{U}
\end{eqnarray}

By using the operator $\hat{U}$, we could assign different values of 
U(1) charges to the each local coordinate.
Furthermore, different from method of Ref.\cite{seki1}, we could indicate 
the necessity of the restrictions on the superpotential explicitly, and 
succeeded in constructing a more general (0,2) Lagrangian density, which 
has different U(1) charges for each local coordinate.

\section{$\left(0,2\right)$ Supermultiplets in the Non-Abelian Gauge Theory}
Now we will construct the (0,2) Lagrangian density for the U(N) gauge group.
In contrast to the U(1) case, in the U(N) case we do not need to assign
different values of the charge to each local coordinate.
By introducing the (0,2) supermultiplets in the U(N) gauge group, we can 
construct the (0,2) U(N) Lagrangian density without using the $\hat{U}$ operator at first.

First, $\Psi=\theta^{+}\overline{\theta}^{+}\sum_{a}v_{+}^{a}T^{a}$ is 
defined for the vector fields $v_{\mu}$, where $T^{a}$ are the generators 
of the ${\rm U(N)}$ group and $a=1,\cdots,\rm{dim}$ ${\rm U(N)}$.
In $D=2$, the $(2,2)$ Grassmann even chiral superfield $\Phi_{\left(2,2\right)i}$ and $(2,2)$ Grassmann odd
chiral superfield $\Xi_{\left(2,2\right)i}$ are given in a manner similar to the ${\rm U(1)}$ case:
\begin{eqnarray}
&&\hspace{-0.5cm}\Phi^{I}_{\left(2,2\right)i}
\nonumber\\
&&\hspace{-0.9cm}=\phi^{I}_{i}
+\sqrt{\mathstrut 2}\left(\theta^{+}\psi_{+i}^{I}
+\theta^{-}\psi_{-i}^{I}\right)
+2\theta^{+}\theta^{-}F_{i}^{I}
-i\theta^{-}\overline{\theta}^{-}\partial_{-}\phi_{i}^{I}
-i\theta^{+}\overline{\theta}^{+}\partial_{+}\phi_{i}^{I}
\nonumber\\
&&\hspace{-0.5cm}-\sqrt{\mathstrut 2}i\theta^{+}\theta^{-}\overline{\theta}^{-}\partial_{-}\psi_{+i}^{I}
+\sqrt{\mathstrut 2}i\theta^{+}\theta^{-}\overline{\theta}^{+}\partial_{+}\psi_{-i}^{I}
-\theta^{+}\theta^{-}\overline{\theta}^{-}\overline{\theta}^{+}\partial_{-}\partial_{+}\phi_{i}^{I},
\\
\nonumber\\
&&\hspace{-0.5cm}\Xi^{A}_{\left(2,2\right)i}
\nonumber\\
&&\hspace{-0.9cm}
=\xi_{i}^{A}
+\sqrt{\mathstrut 2}\left(\theta^{+}b_{+i}^{A}
+\theta^{-}b_{-i}^{A}\right)
+2\theta^{+}\theta^{-}\chi_{i}^{A}
-i\theta^{-}\overline{\theta}^{-}\partial_{-}\xi_{i}^{A}
-i\theta^{+}\overline{\theta}^{+}\partial_{+}\xi_{i}^{A}
\nonumber\\
&&\hspace{-0.5cm}-\sqrt{\mathstrut 2}i\theta^{+}\theta^{-}\overline{\theta}^{-}\partial_{-}b_{+i}^{A}
+\sqrt{\mathstrut 2}i\theta^{+}\theta^{-}\overline{\theta}^{+}\partial_{+}b_{-i}^{A}
-\theta^{+}\theta^{-}\overline{\theta}^{-}\overline{\theta}^{+}\partial_{-}\partial_{+}\xi_{i}^{A},
\end{eqnarray}
where $i=1,\cdots,N$ \cite{seki1}.
For these $(2,2)$ chiral superfields, we will define superfields with restrictions $\theta^{-}=\overline{\theta}^{-}=0$ as follows:
\begin{eqnarray}
&&\Phi_{\left(0,2\right)i}^{I}\equiv\sum_{j}\left(e^\Psi\right)_{ij}\Phi_{\left(2,2\right)j}^{I}\Biggl|_{\theta^{-}=\overline{\theta}^{-}=0},
\label{Phi02non.}\\
&&\Xi_{\left(0,2\right)i}^{A}\equiv\sum_{j}\left(e^{\Psi}\right)_{ij}\Xi_{\left(2,2\right)j}^{A}\Biggl|_{\theta^{-}=\overline{\theta}^{-}=0},
\label{Xi02non.}
\end{eqnarray}
where the covariant derivatives of the gauge transformation for the 
components of the $(2,2)$ chiral superfields are defined as:
\begin{eqnarray}
&&\left(D_{\mu}\phi^{I}\right)_{i}=\partial_{\mu}\phi_{i}^{I}+i\sum_{j}v_{\mu ij}\phi_{j}^{I}.
\end{eqnarray}

We now consider the Lagrangian density $\mathcal{L}_{non.kin.}$ for the 
fields in Eqs. (\ref{Phi02non.}) and (\ref{Xi02non.}). 
From the definition of the $\left(2,2\right)$ vector superfield, 
\begin{eqnarray}
&&\hspace{-0.5cm}\mathcal{V}=\sum_{a}\left(iv_{-}^a+2\theta^{+}\overline{\lambda}_{-}^a+2\overline{\theta}^{+}\lambda_{-}^a+2i\theta^{+}\overline{\theta}^{+}D^a\right)T^a,
\end{eqnarray}
we can define the covariant derivative:
\begin{eqnarray}
\mathcal{D}_{0}-\mathcal{D}_{3}\equiv\partial_{-}+\mathcal{V}.
\label{mathcalD0-mathcalD3non.}
\end{eqnarray}
Then, $\mathcal{L}_{non.kin.}$ is given by Eqs. (\ref{Phi02non.}), (\ref{Xi02non.}) and (\ref{mathcalD0-mathcalD3non.})
\begin{eqnarray}
&&\mathcal{L}_{non.kin.}=\frac{i}{2}\int d\theta^{+}d\overline{\theta}^{+}\sum_{i,j}\Biggl[\sum_{I}\overline{\Phi}_{\left(0,2\right)i}^{I}\left(\mathcal{D}_{0}-\mathcal{D}_{3}\right)_{ij}\Phi_{\left(0,2\right)j}^{I}
\nonumber\\
&&\hspace{4.7cm}+\sum_{A}\overline{\Xi}_{\left(0,2\right)i}^{A}\left(\mathcal{D}_{0}-\mathcal{D}_{3}\right)_{ij}\Xi_{\left(0,2\right)j}^{A}\Biggr].
\label{mathcalLnon.kin.}
\end{eqnarray}

The Lagrangian density $\mathcal{L}_{non.gauge}$ for the vector superfield $\mathcal{V}$ and Fayet-Iliopoulos(FI) term $\mathcal{L}_{non.D,\theta}$, 
which arises from ${\rm U(1)}$ sector of the ${\rm U(N)}$ group, is given as follows.
We define an operator acting on a function $f_{i}(x_{\mu},\theta^{+},\overline{\theta}^{+})$ as
\begin{eqnarray}
&&\sum_{j}\mathcal{D}_{+ij}f_{j}\equiv\sum_{j,k}\left(e^{-\Psi}\right)_{ik}\left(\frac{\partial}{\partial\theta^{+}}-i\overline{\theta}^{+}\partial_{+}\right)\left(e^{\Psi}\right)_{kj}f_{j}.
\label{mathcalD+non.}
\end{eqnarray}
From Eqs. (\ref{mathcalD0-mathcalD3non.}) and (\ref{mathcalD+non.}), we obtain
\begin{eqnarray}
&&\Upsilon_{non.}\equiv\bigl[\mathcal{\overline{D}}_{+},\left(\mathcal{D}_{0}-\mathcal{D}_{3}\right)\bigr].
\label{Upsilonnon.}
\end{eqnarray}
The covariant derivatives of the gauge transformations for the components 
of the $(2,2)$ vector superfield are given by
\begin{eqnarray}
&&D_{\pm}\lambda_{-}=\partial_{\pm}\lambda_{-}+i\bigl[v_{\pm},\lambda_{-}\bigr].
\end{eqnarray}
From Eq. (\ref{Upsilonnon.}), $\mathcal{L}_{non.gauge}$ can be given as
\begin{eqnarray}
&&\hspace{-1cm}\mathcal{L}_{non.gauge}=-\frac{1}{8e^{2}}\int d\theta^{+}d\overline{\theta}^{+}{\rm tr}\Biggl[\Upsilon_{non.}\overline{\Upsilon}_{non.}\Biggr],
\label{mathcalLnon.gauge}
\end{eqnarray}
and $\mathcal{L}_{non.D,\theta}$ is
\begin{eqnarray}
&&\mathcal{L}_{non.D,\theta}=\frac{t}{4}\int d\theta^{+}{\rm tr}\Upsilon_{non.}\Bigl|_{\overline{\theta}^{+}=0}+\frac{\overline{t}}{4}\int d\overline{\theta}^{+}{\rm tr}\overline{\Upsilon}_{non.}\Bigr|_{\theta^{+}=0}.
\label{mathcalLnon.D,theta}
\end{eqnarray}
Since Eq. (\ref{Upsilonnon.}) includes only part of the components of the 
$(2,2)$ vector superfield, the residual compensating components will be given by the superfield $\Omega$
:
\begin{eqnarray}
&&\Omega\equiv\sum_{a}\left(\sigma^a+\sqrt{\mathstrut 2}i\theta^{+}\overline{\lambda}_{+}^a-i\theta^{+}\overline{\theta}^{+}\partial_{+}\sigma^a\right)T^{a}.
\label{Omeganon.''}
\end{eqnarray}
From Eq. (\ref{Omeganon.''}), we redefine the following superfield:
\begin{eqnarray}
&&\Omega_{non.}\equiv\Omega+\bigl[\Psi,\Omega\bigr],
\label{Omeganon.}
\end{eqnarray}
and from these definitions, we obtain the following:
\begin{eqnarray}
&&\mathcal{V}^{'}\equiv\sum_{a}\left(iv_{-}^a+2\sqrt{\mathstrut 2}\theta^{+}\overline{\lambda}_{-}^a+2\sqrt{\mathstrut 2}\overline{\theta}^{+}\lambda_{-}^a+2i\theta^{+}\overline{\theta}^{+}D^a\right)T^{a},
\label{mathcalV'non.}\\
&&\Omega^{'}\equiv
\sum_{a}\left(\sigma^a+i\theta^{+}\overline{\lambda}_{+}^a-i\theta^{+}\overline{\theta}^{+}\partial_{+}\sigma^a\right)T^{a}.
\label{Omega'non.}
\end{eqnarray}
Using Eq. (\ref{Omega'non.}), we can define 
\begin{eqnarray}
&&\Omega_{non.}^{'}\equiv\Omega^{'}+\bigl[\Psi,\Omega^{'}\bigr].
\label{Omeganon.'}
\end{eqnarray}
From Eqs. (\ref{mathcalV'non.}) and (\ref{Omeganon.'}), 
\begin{eqnarray}
&&\Gamma\equiv\bigl[\mathcal{V}^{'},\Omega_{non.}^{'}\bigr]\Bigl|_{\overline{\theta}^{+}=0},
\label{Gammanon.}
\end{eqnarray}
is defined.
Then, the Lagrangian density $\mathcal{L}_{non.gauge}^{'}$ is obtained from Eqs. (\ref{Omeganon.}), (\ref{Omeganon.'}) and (\ref{Gammanon.}):
\begin{eqnarray}
&&\mathcal{L}_{non.gauge}^{'}
=\frac{i}{2e^{2}}\int d\theta^{+}d\overline{\theta}^{+}{\rm tr}\Biggl[\overline{\Omega}_{non.}\partial_{-}\Omega_{non.}+\overline{\Omega}_{non.}^{'}\Gamma-\overline{\Gamma}\Omega_{non.}^{'}
\nonumber\\
&&\hspace{5cm}-i\theta^{+}\overline{\theta}^{+}\bigl[\Omega_{non.}^{'},\overline{\Omega}_{non.}^{'}\bigr]^{2}\Biggr].
\label{mathcalLnon.gauge'}
\end{eqnarray}

Next, we will introduce the other $(0,2)$ superfields as follows:
\begin{eqnarray}
&&\Lambda_{Ii}^{'}\equiv\psi_{-i}^{I}-\sqrt{\mathstrut 2}\theta^{+}F_{i}^{I}-i\theta^{+}\overline{\theta}^{+}\partial_{+}\psi_{-i}^{I},
\label{Lambda'non.}\\
&&\tilde{\Lambda}_{Ai}^{'}\equiv b_{-i}^{A}-\sqrt{\mathstrut 2}\theta^{+}\chi_{i}^{A}-i\theta^{+}\overline{\theta}^{+}\partial_{+}b_{-i}^{A}.
\label{tildeLambda'non.}
\end{eqnarray}
We will give functions $E_{Ii}(\Phi_{\left(0,2\right)},\Omega)$, 
and $\tilde{E}_{Ai}(\Xi_{\left(0,2\right)},\Omega)$ 
defined on the variables given by Eqs. (\ref{Phi02non.}), (\ref{Xi02non.}) and (\ref{Omeganon.''}).
We assume these functions to be separable in variables:
\begin{eqnarray}
&&E_{Ii}(\Phi_{\left(0,2\right)},\Omega)=\sum_{j}H_{ij}(\Omega)G_{Ij}(\Phi_{\left(0,2\right)}),
\label{Enon.}\\
&&\tilde{E}_{Ai}(\Xi_{\left(0,2\right)},\Omega)=\sum_{j}H_{ij}(\Omega)\tilde{G}_{Aj}(\Xi_{\left(0,2\right)}).
\label{tildeEnon.}
\end{eqnarray}
Using Eqs. (\ref{Lambda'non.})-(\ref{tildeEnon.}), we redefine the fields:
\begin{eqnarray}
&&\Lambda_{Ii}\equiv\Lambda_{Ii}^{'}-\sqrt{\mathstrut 2}\overline{\theta}^{+}E_{Ii}(\Phi_{\left(0,2\right)},\Omega),
\label{Lambdanon.}
\\
&&\tilde{\Lambda}_{Ai}\equiv\tilde{\Lambda}_{Ai}^{'}-\sqrt{\mathstrut 2}\overline{\theta}^{+}\tilde{E}_{Ai}(\Xi_{\left(0,2\right)},\Omega).
\label{tildeLambdanon.}
\end{eqnarray}
We can then obtain the following identities from Eqs. (\ref{Enon.})-(\ref{tildeLambdanon.}):
\begin{eqnarray}
&&\hspace{-0.5cm}\Lambda_{Ii\left(0,2\right)}\equiv\sum_{j}\left(e^{\Psi}\right)_{ij}\Lambda_{Ij},
\label{Lambda02non.}\\
&&\hspace{-0.5cm}\tilde{\Lambda}_{Ai\left(0,2\right)}\equiv\sum_{j}\left(e^{\Psi}\right)_{ij}\tilde{\Lambda}_{Aj},
\label{tildeLambda02non.}
\end{eqnarray}
and
\begin{eqnarray}
&&E_{Ii}^{'}(\Phi_{\left(0,2\right)}\overline{\Omega})=\sum_{j}\overline{H}_{ij}(\overline{\Omega})G_{Ij}(\Phi_{\left(0,2\right)}),
\label{E'non.}
\\
&&\tilde{E}_{Ai}^{'}(\Xi_{\left(0,2\right)},\overline{\Omega})=\sum_{j}\overline{H}_{ij}(\overline{\Omega})\tilde{G}_{Aj}(\Xi_{\left(0,2\right)}).
\label{tildeE'non.}
\end{eqnarray}

We can now obtain the Lagrangian density $\mathcal{L}_{non.\Lambda}$ from Eqs. (\ref{Enon.}), (\ref{tildeEnon.}), and 
(\ref{Lambda02non.})-(\ref{tildeE'non.}):
\begin{eqnarray}
&&\hspace{-0.35cm}\mathcal{L}_{non.\Lambda}
\nonumber\\
&&\hspace{-0.75cm}=\frac{1}{2}\int d\theta^{+}d\overline{\theta}^{+}\sum_{i}\Biggl[\sum_{I}\overline{\Lambda}_{Ii\left(0,2\right)}\Lambda_{Ii\left(0,2\right)}+\sum_{A}\overline{\tilde{\Lambda}}_{Ai\left(0,2\right)}\tilde{\Lambda}_{Ai\left(0,2\right)}
\nonumber\\
&&\hspace{2.45cm}+\sum_{I}\left(\overline{\theta}^{+}\overline{E}_{Ii}(\overline{\Phi}_{\left(0,2\right)},\overline{\Omega})+\theta^{+}\overline{E}_{Ii}^{'}(\overline{\Phi}_{\left(0,2\right)},\Omega)\right)
\nonumber\\
&&\hspace{2.45cm}\times\left(\theta^{+}E_{Ii}(\Phi_{\left(0,2\right)},\Omega)+\overline{\theta}^{+}E_{Ii}^{'}(\Phi_{\left(0,2\right)},\overline{\Omega})\right)
\nonumber\\
&&\hspace{2.45cm}-\sum_{A}\left(\overline{\theta}^{+}\overline{\tilde{E}}_{Ai}(\overline{\Xi}_{\left(0,2\right)},\overline{\Omega})+\theta^{+}\overline{\tilde{E}}_{Ai}^{'}(\overline{\Xi}_{\left(0,2\right)},\Omega)\right)
\nonumber\\
&&\hspace{2.45cm}\times\left(\theta^{+}\tilde{E}_{Ai}(\Xi_{\left(0,2\right)},\Omega)+\overline{\theta}^{+}\tilde{E}_{Ai}^{'}(\Xi_{\left(0,2\right)},\overline{\Omega})\right)\Biggr].
\label{mathcalLnon.Lambda}
\end{eqnarray}
We will choose the functions given in Eqs. (\ref{Enon.}) and (\ref{tildeEnon.}) as
\begin{eqnarray}
\begin{array}{rcl}
&&G_{Ii}(\phi)=\sqrt{\mathstrut 2}\phi_{i}^{I},
\\
&&\tilde{G}_{Ai}(\xi)=\sqrt{\mathstrut 2}\xi_{i}^{A},
\\
&&H(\sigma)=\sigma.
\end{array}
\label{N=left(0,2right)non.}
\end{eqnarray}
\label{mathcalLnon.Lambda'}
We further define functions $J_{i}^{I}(\Phi_{\left(0,2\right)},\Xi_{\left(0,2\right)})$ 
and 
$\tilde{J}_{i}^{A}(\Phi_{\left(0,2\right)},\Xi_{\left(0,2\right)})$
 by Eqs. (\ref{Phi02non.}) and (\ref{Xi02non.}),
and assume the following relations:
\begin{eqnarray}
\begin{array}{rcl}
&&J_{i}^{I}(\phi,\xi)=\frac{\partial W(\phi,\xi)}{\partial\phi_{i}^{I}},
\\
&&\tilde{J}_{i}^{A}(\phi,\xi)=\frac{\partial W(\phi,\xi)}{\partial\xi_{i}^{A}},
\end{array}
\label{Jnon.}
\end{eqnarray}
where $W$ is superpotential of the theory.
The Lagrangian density $\mathcal{L}_{non.J}$ is then obtained by Eqs. (\ref{Lambda02non.}), (\ref{tildeLambda02non.}) and (\ref{Jnon.}) as follows:
\begin{eqnarray}
&&\hspace{-0.1cm}\mathcal{L}_{non.J}
\nonumber\\
&&\hspace{-0.5cm}=\frac{1}{\sqrt{\mathstrut 2}}\int d\theta^{+}\sum_{i}\Biggl[\sum_{I}\Lambda_{Ii\left(0,2\right)}J_{i}^{I}(\Phi_{\left(0,2\right)},\Xi_{\left(0,2\right)})
\nonumber\\
&&\hspace{2.5cm}+\sum_{A}\tilde{\Lambda}_{Ai\left(0,2\right)}\tilde{J}_{i}^{A}(\Phi_{\left(0,2\right)},\Xi_{\left(0,2\right)})\Biggr]\Biggl|_{\overline{\theta}^{+}=0}
\nonumber\\
&&\hspace{-0.1cm}+\frac{1}{\sqrt{\mathstrut 2}}\int d\overline{\theta}^{+}\sum_{i}\Biggl[\sum_{I}\overline{J}_{i}^{I}(\overline{\Phi}_{\left(0,2\right)},\overline{\Xi}_{\left(0,2\right)})\overline{\Lambda}_{Ii\left(0,2\right)}
\nonumber\\
&&\hspace{2.8cm}+\sum_{A}\overline{\tilde{J}}_{i}^{A}(\overline{\Phi}_{\left(0,2\right)},\overline{\Xi}_{\left(0,2\right)})\overline{\tilde{\Lambda}}_{Ai\left(0,2\right)}\Biggr]\Biggr|_{\theta^{+}=0}.
\label{mathcalLnon.J}
\end{eqnarray}
Finally, by combining Eqs. (\ref{mathcalLnon.kin.}), (\ref{mathcalLnon.gauge}), (\ref{mathcalLnon.D,theta}), 
(\ref{mathcalLnon.gauge'}), (\ref{mathcalLnon.Lambda}), and
(\ref{mathcalLnon.J}), we can obtain the $(0,2)$ total Lagrangian density $\mathcal{L}_{\left(0,2\right)non.}$:  
\begin{eqnarray}
&&\mathcal{L}_{\left(0,2\right)non.}
=\mathcal{L}_{non.kin.}
+\mathcal{L}_{non.gauge}
+\mathcal{L}_{non.D,\theta}
+\mathcal{L}_{non.gauge}^{'}
\nonumber\\
&&\hspace{1.9cm}+\mathcal{L}_{non.\Lambda}
+\mathcal{L}_{non.J}.
\label{l}
\end{eqnarray}
In Eq. (\ref{l}), the (0,2) U(N) Lagrangian density was constructed 
without using $\hat{U}$ at this moment, because we do not need to 
assign differenent values of charge to each local coordinate.

\section{$(0,2)$ Supersymmetric Transformation and Invariance of Lagrangian Densities in Non-Abelian Gauge Theory}
In this section, we will verify the $\left(0,2\right)$ supersymmetric transformation properties of Eqs. (\ref{mathcalLnon.kin.}), (\ref{mathcalLnon.gauge}), 
(\ref{mathcalLnon.D,theta}), (\ref{mathcalLnon.gauge'}), (\ref{mathcalLnon.Lambda}), and (\ref{mathcalLnon.J}).

In constructing the (0,2) U(N) Lagrangian density, it appears that 
similar restrictions on the superpotential are required as in the U(1) case.
So here we concentrate on the restrictions on the superpotential, and 
compare them for the U(N) and the U(1) cases.
While for the U(1) part, a Calabi-Yau supermanifold with the same number of 
even coordinates and odd coordinates may be obtained, for the SU(N) part 
constraints, a supermanifold may be defined which is different from the 
U(1) case.

The $\left(0,2\right)$ supersymmetric transformation properties of the 
components of the $(2,2)$ Grassmann even and odd superfields are given, 
respectively, as follows:
\begin{eqnarray}
\hspace{-1cm}\begin{array}{rcl}
&&\delta_{\left(0,2\right)}\phi_{i}^{I}=-\sqrt{\mathstrut 2}\epsilon_{-}\psi_{+i}^{I},
\\
&&\delta_{\left(0,2\right)}\psi_{+i}^{I}=\sqrt{\mathstrut 2}i\overline{\epsilon}_{-}\left(D_{+}\phi^{I}\right)_{i},
\\

&&\delta_{\left(0,2\right)}\psi_{-i}^{I}=\sqrt{\mathstrut 2}\epsilon_{-}F_{i}^{I}+2\overline{\epsilon}_{-}\sum_{j}\sigma_{ij}\phi_{j}^{I},
\\
&&\delta_{\left(0,2\right)}F_{i}^{I}=\overline{\epsilon}_{-}\Bigg\{-\sqrt{\mathstrut 2}i\left(D_{+}\psi_{-}^{I}\right)_{i}+2\sum_{j}\sigma_{ij}\psi_{+j}^{I}+2i\sum_{j}\overline{\lambda}_{+ij}\phi_{j}^{I}\Bigg\},
\end{array}
\label{deltaPhinon.}
\end{eqnarray}

\begin{eqnarray}
\hspace{-1cm}\begin{array}{rcl}
&&\delta_{\left(0,2\right)}\xi_{i}^{A}=-\sqrt{\mathstrut 2}\epsilon_{-}b_{+i}^{A},\\
&&\delta_{\left(0,2\right)}b_{+i}^{A}=\sqrt{\mathstrut 2}i\overline{\epsilon}_{-}\left(D_{+}\xi^{A}\right)_{i},
\\
&&\delta_{\left(0,2\right)}b_{-i}^{A}=\sqrt{\mathstrut 2}\epsilon_{-}\chi_{i}^{A}+2\overline{\epsilon}_{-}\sum_{j}\sigma_{ij}\xi_{j}^{A}
,\\
&&\delta_{\left(0,2\right)}\chi_{i}^{A}=\overline{\epsilon}_{-}\Bigg\{-\sqrt{\mathstrut 2}i\left(D_{+}b_{-}^{A}\right)_{i}+2\sum_{j}\sigma_{ij}b_{+j}^{A}+2i\sum_{j}\overline{\lambda}_{+ij}\xi_{j}^{A}\Bigg\}.
\end{array}
\label{deltaXinon.}
\end{eqnarray}

The $\left(0,2\right)$ supersymmetric transformation properties of components 
of the $(2,2)$ vector superfield are given as:
\begin{eqnarray}
\hspace{-1cm}\begin{array}{rcl}
&&\delta_{\left(0,2\right)}v_{+}=0
,\\
&&\delta_{\left(0,2\right)}v_{-}=2i\left(\epsilon_{-}\overline{\lambda}_{-}+\overline{\epsilon}_{-}\lambda_{-}\right)
,\\
&&\delta_{\left(0,2\right)}\sigma=-\sqrt{\mathstrut 2}i\epsilon_{-}\overline{\lambda}_{+}
,\\
&&\delta_{\left(0,2\right)}\lambda_{-}=i\epsilon_{-}\left(D-iv_{03non.}-\bigl[\sigma,\overline{\sigma}\bigr]\right)
,\\
&&\delta_{\left(0,2\right)}\lambda_{+}=\sqrt{\mathstrut 2}\epsilon_{-}D_{+}\overline{\sigma}
,\\
&&\delta_{\left(0,2\right)}D=\epsilon_{-}\left(D_{+}\overline{\lambda}_{-}+\sqrt{\mathstrut 2}i\bigl[\overline{\sigma},\overline{\lambda}_{+}\bigr]\right)-\overline{\epsilon}_{-}\left(D_{+}\lambda_{-}+\sqrt{\mathstrut 2}i\bigl[\sigma,\lambda_{+}\bigr]\right)
,
\end{array}
\label{deltaVnon.}
\end{eqnarray}
where $v_{03non.}=\partial_{0}v_{3}-\partial_{3}v_{0}+i\bigl[v_{0},v_{3}\bigr]$.

Using the Eqs. (\ref{deltaPhinon.})-(\ref{deltaVnon.}), the actions for $\mathcal{L}_{\left(0,2\right)non.}$ are supersymmetric invariants up to total derivatives.
However, for ${\cal L}_{non.J}$ in Eq. (\ref{mathcalLnon.J}), we obtain:
\begin{eqnarray}
&&\hspace{-0.5cm}\delta_{\left(0,2\right)}\mathcal{L}_{non.J}
\nonumber\\
&&\hspace{-0.9cm}=\sqrt{\mathstrut 2}i\overline{\epsilon}_{-}\partial_{+}\sum_{i}\Biggl[\sum_{I}\psi_{-i}^{I}\frac{\partial W(\phi,\xi)}{\partial\phi_{i}^{I}}
+\sum_{A}b_{-i}^{A}\frac{\partial W(\phi,\xi)}{\partial\xi_{i}^{A}}\Biggr]
\nonumber\\
&&\hspace{-0.5cm}-\sqrt{\mathstrut 2}\overline{\epsilon}_{-}\sum_{i,j}\Biggl[\sum_{k,J}v_{+ij}\psi_{-k}^{J}\frac{\partial}{\partial\phi_{k}^{J}}
+\sqrt{\mathstrut 2}\sum_{k,J}\sigma_{ij}\psi_{+k}^{J}\frac{\partial}{\partial\phi_{k}^{J}}
+\sum_{k,B}v_{+ij}b_{-k}^{B}\frac{\partial}{\partial\xi_{k}^{B}}
\nonumber\\
&&\hspace{-0.5cm}+\sqrt{\mathstrut 2}\sum_{k,B}\sigma_{ij}b_{+k}^{B}\frac{\partial}{\partial\xi_{k}^{B}}
+\sqrt{\mathstrut 2}i\overline{\lambda}_{+ij}\Biggr]
\Biggl[\sum_{I}\phi_{j}^{I}\frac{\partial W(\phi,\xi)}{\partial\phi_{i}^{I}}+\sum_{A}\xi^{A}_{j}\frac{\partial W(\phi,\xi)}{\partial\xi_{i}^{A}}\Biggr]
+(h.c.).
\label{delta02mathcalLnon.J}
\end{eqnarray}
Next, we derive the consistency condition for the $(0,2)$ supersymmetric invariances of the action under the $(0,2)$ 
supersymmetric transformation by the following relation using Eq. (\ref{delta02mathcalLnon.J}):
\begin{eqnarray}
&&\sum_{i,j}\left(\sum_{I}T_{ij}^{a}\phi_{j}^{I}\frac{\partial W(\phi,\xi)}{\partial\phi_{i}^{I}}
+\sum_{A}T_{ij}^{a}\xi_{j}^{A}\frac{\partial W(\phi,\xi)}{\partial\xi_{i}^{A}}\right)=0.
\label{zyoukennon.1}
\end{eqnarray}
We define Eq.(\ref{zyoukennon.1}) as a function $G(\phi,\xi)$:
\begin{eqnarray}
G(\phi,\xi)\equiv\sum_{i,j}\left(\sum_{I}T^{a}_{ij}\phi_{j}^{I}\frac{\partial W(\phi,\xi)}{\partial\phi_{i}^{I}}
+\sum_{A}T^{a}_{ij}\xi_{j}^{A}\frac{\partial W(\phi,\xi)}{\partial\xi_{i}^{A}}\right)=0,
\label{tuika1}
\end{eqnarray}
and transform the function $G(\phi,\xi)$ under the transformation laws:
\begin{eqnarray}
\phi^{I}_{i}\rightarrow\sum_{j}(\lambda^{T^{a}})_{ij}\phi_{j}^{I},\hspace{0.5cm}\xi_{i}^{A}\rightarrow\sum_{j}(\lambda^{T^{a}})_{ij}\xi_{j}^{A}.
\label{tuika2}
\end{eqnarray}
Because $G(\phi,\xi)$ is equal to zero, the function transformed by using Eq.(\ref{tuika2}) also vanishes:
\begin{eqnarray}
G(\phi,\xi)=G(\lambda^{T}\phi,\lambda^{T}\xi)=0.
\label{tuika3}
\end{eqnarray}
Eq.(\ref{tuika3}) gives the equivalence relation for local coordinates in a supermanifold $\mathcal{M}^{m|n}$:
\begin{eqnarray}
&&\hspace{-1.5cm}(\phi_{i}^{1},\cdots,\phi_{i}^{m}|\xi_{i}^{1},\cdots,\xi_{i}^{n})
\nonumber\\
&&\sim(\sum_{j}(\lambda^{T^{a}})_{ij}\phi_{j}^{1},\cdots,\sum_{j}(\lambda^{T^{a}})_{ij}\phi_{j}^{m}\Bigl|\sum_{j}(\lambda^{T^{a}})_{ij}\xi_{j}^{1},\cdots,\sum_{j}(\lambda^{T^{a}})_{ij}\xi_{j}^{n}).
\label{tuika4}
\end{eqnarray}
Eq.(\ref{tuika4}) may indicate that the manifold $\mathcal{M}^{m|n}$ 
can be considered to be equivalent to the super weighted projective space $WCP^{m-1|n}$. 
Moreover, we calculate for the superpotential $W(\lambda^{T}\phi,\lambda^{T}\xi)$ as follows:
\begin{eqnarray}
&&\sum_{i,j}\left(\sum_{I}T^{a}_{ij}\phi_{j}^{I}\frac{\partial W(\lambda^{T}\phi,\lambda^{T}\xi)}{\partial\phi_{i}^{I}}
+\sum_{A}T^{a}_{ij}\xi_{j}^{A}\frac{\partial W(\lambda^{T}\phi,\lambda^{T}\xi)}{\partial\xi_{i}^{A}}\right)
\nonumber\\
&&\hspace{-0.5cm}=\sum_{b,i,j,k,m}\left(\sum_{I}T^{a}_{ij}\phi_{j}^{I}(\lambda^{T^{b}})_{ki}\frac{\partial W(\lambda^{T}\phi,\lambda^{T}\xi)}{\partial((\lambda^{T^{b}})_{km}\phi_{m}^{I})}
+\sum_{A}T^{a}_{ij}\xi_{j}^{A}(\lambda^{T^{b}})_{ki}\frac{\partial W(\lambda^{T}\phi,\lambda^{T}\xi)}{\partial((\lambda^{T^{b}})_{km}\xi_{m}^{A})}\right).
\label{tuika5}
\end{eqnarray}
Eq.(\ref{tuika5}) can be divided into the U(1) part and the SU(N) part of the U(N) gauge group.
Then, the generators of the U(1) part and the SU(N) part are defined as $T^{0}_{ij}=M\delta_{ij}$ and $T^{\Theta}_{ij}$ $(\Theta=1,\cdots,\rm{N}^{2}-1)$ respectively, where $M$ is a normalization factor.
The U(1) part of Eq.(\ref{tuika5}) is calculated as follows:
\begin{eqnarray}
M\sum_{b,i,j,k}\left(\sum_{I}(\lambda^{T^{b}})_{ij}\phi_{j}^{I}\frac{\partial W(\lambda^{T}\phi,\lambda^{T}\xi)}{\partial((\lambda^{T^{b}})_{ik}\phi_{k}^{I})}
+\sum_{A}(\lambda^{T^{b}})_{ij}\xi_{j}^{A}\frac{\partial W(\lambda^{T}\phi,\lambda^{T}\xi)}{\partial((\lambda^{T^{b}})_{ik}\xi_{k}^{A})}\right),
\label{tuika6}
\end{eqnarray}
which coincides with the U(1) part of $G(\lambda^{T}\phi,\lambda^{T}\xi)$.
Furthermore, by using Eq.(\ref{tuika3}), we found that these equations also 
coincide with that of $G(\phi,\xi)$:
\begin{eqnarray}
&&\hspace{-1.8cm}G(\phi,\xi)\Bigl|_{a=0}=G(\lambda^{T}\phi,\lambda^{T}\xi)\Bigr|_{a=0}
\nonumber\\
&&=M\sum_{i}\left(\sum_{I}\phi_{i}^{I}\frac{\partial W(\phi,\xi)}{\partial\phi_{i}^{I}}
+\sum_{A}\xi_{i}^{A}\frac{\partial W(\phi,\xi)}{\partial\xi_{i}^{A}}\right)
\nonumber\\
&&=M\sum_{i}\left(\sum_{I}\phi_{i}^{I}\frac{\partial W(\lambda^{T}\phi,\lambda^{T}\xi)}{\partial\phi_{i}^{I}}
+\sum_{A}\xi_{i}^{A}\frac{\partial W(\lambda^{T}\phi,\lambda^{T}\xi)}{\partial\xi_{i}^{A}}\right)
\nonumber\\
&&=0.
\label{tuika8}
\end{eqnarray}
Therefore, Eq.(\ref{tuika8}) gives the quasi-homogeneous condition $W(\phi,\xi)=W(\lambda^{T}\phi,\lambda^{T}\xi)$ for the superpotential.

On the other hand, the SU(N) part of Eq.(\ref{tuika5}) is:
\begin{eqnarray}
\sum_{b,i,j,k,l}\left(\sum_{I}(\lambda^{T^{b}})_{ij}T^{\Theta}_{jk}\phi_{k}^{I}\frac{\partial W(\lambda^{T}\phi,\lambda^{T}\xi)}{\partial((\lambda^{T^{b}})_{il}\phi_{l}^{I})}
+\sum_{A}(\lambda^{T^{b}})_{ij}T^{\Theta}_{jk}\xi_{k}^{A}\frac{\partial W(\lambda^{T}\phi,\lambda^{T}\xi)}{\partial((\lambda^{T^{b}})_{il}\xi_{l}^{A})}\right),
\label{tuika7}
\end{eqnarray}
which, however, does not coincide with the SU(N) part of $G(\phi,\xi)$ and of $G(\lambda^{T}\phi,\lambda^{T}\xi)$.
Thus the superpotential $W(\phi,\xi)$ does not satisfy a quasi-homogeneous condition in SU(N).

From these results, the supermanifold $\mathcal{M}^{m|n}$ seems to become the super weighted complex projective space $WCP^{m-1|n}$, although
 the superpotential $W(\phi,\xi)$ for the non-Abelian gauge group does not satisfy a quasi-homogeneous condition in SU(N), except for the U(1) part of U(N).
Therefore, because of the extention to the U(N) gauge group, there are more
stringent restrictions to be imposed on the form of the 
superpotential than in the U(1) case.

From the U(1) part, the Calabi-Yau supermanifold must have the same number 
of even coordinates and odd coordinates from Eq. (\ref{q1}).
In the SU(N) part, we must take care in constructing the Calabi-Yau 
supermanifold, because there are more stringent restrictions to be imposed on
the form of the superpotential than in the U(1) case.

\section{${\rm U(N)}$ Charge Operator}
In constructing the (0,2) U(N) Lagrangian density, we could not confirm the 
reason of necessity to introduce the $\hat{U}$-type operator.
However, in order to introduce the (0,2) chiral superfields, we need this 
operator, because otherwise we cannot define the (0,2) chirality conditions 
of the (0,2) chiral superfieldsm as will be shown later.
Therefore, in this section, we will introduce the $\hat{U}^a$  operator in 
U(N) version in order to define the (0,2) chirality conditions of the 
(0,2) chiral superfields.

We can define the ${\rm U(N)}$ charge operator in a manner similar to that for the ${\rm U(1)}$ Abelian case:
\begin{eqnarray}
&&\hspace{-1cm}\hat{U}_{ij}^{a}
\equiv\frac{1}{N}\sum_{k,I}\Biggl[T_{ik}^{a}\phi_{k}^{I}\frac{\partial}{\partial\phi_{j}^{I}}
+\sum_{\mu}T_{ik}^{a}\partial_{\mu}\phi_{k}^{I}\frac{\partial}{\partial\left(\partial_{\mu}\phi_{j}^{I}\right)}
+\sum_{\mu,\nu}T_{ik}^{a}\partial_{\mu}\partial^{\mu}\phi_{k}^{I}\frac{\partial}{\partial\left(\partial_{\nu}\partial^{\nu}\phi_{j}^{I}\right)}
\nonumber\\
&&+\sum_{\alpha}T_{ik}^{a}\psi_{\alpha k}^{I}\frac{\partial}{\partial\psi_{\alpha j}^{I}}
+\sum_{\mu,\alpha}T_{ik}^{a}\partial_{\mu}\psi_{\alpha k}^{I}\frac{\partial}{\partial\left(\partial_{\mu}\psi_{\alpha j}^{I}\right)}
+T_{ik}^{a}F_{k}^{I}\frac{\partial}{\partial F_{j}^{I}}
\Biggr]
\nonumber\\
&&+\frac{1}{N}\sum_{k,A}\Biggl[T_{ik}^{a}\xi_{k}^{A}\frac{\partial}{\partial\xi_{j}^{A}}
+\sum_{\mu}T_{ik}^{a}\partial_{\mu}\xi_{k}^{A}\frac{\partial}{\partial\left(\partial_{\mu}\xi_{j}^{A}\right)}
+\sum_{\mu,\nu}T_{ik}^{a}\partial_{\mu}\partial^{\mu}\xi_{k}^{A}\frac{\partial}{\partial\left(\partial_{\nu}\partial^{\nu}\xi_{j}^{A}\right)}
\nonumber\\
&&+\sum_{\alpha}T_{ik}^{a}b_{\alpha k}^{A}\frac{\partial}{\partial b_{\alpha j}^{A}}
+\sum_{\mu,\alpha}T_{ik}^{a}\partial_{\mu}b_{\alpha k}^{A}\frac{\partial}{\partial\left(\partial_{\mu}b_{\alpha j}^{A}\right)}
+T_{ik}^{a}\chi_{k}^{A}\frac{\partial}{\partial\chi_{j}^{A}}
\Biggr]+(h.c.).
\label{hatUnon.}
\end{eqnarray}
From Eq. (\ref{hatUnon.}), the consistency condition in Eq. (\ref{zyoukennon.1}) is rewritten as:
\begin{eqnarray}
&&N\sum_{i,j}\delta_{ij}\hat{U}_{ij}^{a}W(\phi,\xi)=0.
\end{eqnarray}

Using the operator in Eq. (\ref{hatUnon.}), we are able to define an 
operation on the function $f_{i}(x_{\mu},\theta^{+},\overline{\theta}^{+})$ 
as follows:
\begin{eqnarray}
&&\hspace{-0.8cm}\sum_{j}\mathcal{D}_{+ij}^{'}f_{j}\equiv\sum_{j,k}\left(e^{-\Psi^{'}}\right)_{ik}\left(\frac{\partial}{\partial\theta^{+}}-i\overline{\theta}^{+}\partial_{+}\right)\left(e^{\Psi^{'}}\right)_{kj}f_{j},
\label{mathcalD+'non.}
\end{eqnarray}
where $\Psi^{'}=\theta^{+}\overline{\theta}^{+}\sum_{a}v_{+}^{a}\hat{U}^{a}$ is assumed.
We finally obtain the $\left(0,2\right)$ chirality conditions by using Eqs. (\ref{Phi02non.}), (\ref{Xi02non.}) and (\ref{mathcalD+'non.}):
\begin{eqnarray}
&&\sum_{j}\overline{\mathcal{D}}_{+ij}^{'}\Phi_{\left(0,2\right)j}^{I}=\sum_{i}\mathcal{D}_{+ij}^{'}\overline{\Phi}_{\left(0,2\right)i}^{I}=0,
\\
&&\sum_{j}\overline{\mathcal{D}}_{+ij}^{'}\Xi_{\left(0,2\right)j}^{A}=\sum_{i}\mathcal{D}_{+ij}^{'}\overline{\Xi}_{\left(0,2\right)i}^{A}=0.
\end{eqnarray}
From these results, we could confirm the necessity of the $\hat{U}$ operator for defining the (0,2) chirality conditions of (0,2) chiral superfields, though 
this operator was not required for the construction of the (0,2) U(N) 
Lagrangian density.

\section{Summary and Discussion}

We have constructed the $D=2$, $(0,2)$ U(1) gauged linear sigma model on a 
supermanifold $\mathcal{M}^{m|n}$ by a method which differs from that 
of Ref.\cite{seki1}, because to our opinion that method seems to be 
incomplete.
Furthermore, we have constructed the U(N) gauged linear sigma model explicitly.

In the first part of the present paper, we consistently constructed the $D=2$, $(0,2)$ U(1) gauged linear sigma model on the supermanifold $\mathcal{M}^{m|n}$, by introducing a new operator, $\hat{U}$. 
In the method of Ref.\cite{seki1}, it was impossible to assign different 
value of U(1) charge to each local coordinate.
The explicit form of the $\hat{U}$ operator was determined by assuming that 
it is the operator that assigns different value of U(1) charges to each 
local coordinate of $\mathcal{M}^{m|n}$. 
The covariant derivatives and super charges of the (0,2) supersymmetric 
transformation are also defined using the $\hat{U}$ operator. The (0,2) 
chirality conditions on the of the superpotential term in the Lagrangian 
density appear to be most appropriately implied by these covariant derivatives.

The (0,2) supersymmetric invariance of the Lagrangian density of the (0,2) 
U(1) gauged linear sigma model was also proved by using consistency 
conditions derived by using the $\hat{U}$ operator. 
We found that the conditions that assure the (0,2) supersymmetric invariance of the Lagrangian density agree with the (0,2) chirality conditions for the superpotential. 
Though the method of Ref.\cite{seki1} could not confirm
the necessity of ristriction conditions clearly, 
we could indicate the necessity of conditions explicitly.
The supermanifold $\mathcal{M}^{m|n}$ then becomes the super weighted complex projective space $WCP^{m-1|n}$ from these conditions.
If we focus on the Calabi-Yau supermanifold corresponding to the super 
Landau-Ginzburg model, by using $\hat{U}$ we can construct a Calabi-Yau 
supermanifold which is more
general than in Ref.\cite{seki1}, and which has a different number of even 
coordinates and odd coordinates. 

In the second part of the present paper, we constructed a $D=2$, (0,2) 
U(N) gauged linear sigma model on the supermanifold $\mathcal{M}^{m|n}$ as a new construction.
The construction is approximately parallel to the U(1) case, 
but the $\hat{U}^{a}$ operator, which is an extension of the $\hat{U}$ 
operator of the U(1) gauge group to the U(N) gauge group, coincides with 
a set of generators of U(N). 
Although $\hat{U}^{a}$ is unnecessary in constructing the Lagrangian 
density of the (0,2) U(N)
gauged linear sigma model, we could confirm the necessity of $\hat{U}^{a}$ 
for giving the (0,2) chirality conditions of the (0,2) chiral superfields.
We obtained the conditions that give 
(0,2) supersymmetric invariance of the Lagrangian density of the (0,2) U(N) gauged linear sigma model.

As in the case of U(1), these conditions decide the form of the 
superpotential.
However, in the U(N) case more stringent restrictions on the form of the 
superpotential have to be imposed  than in the U(1) case.
From these results, the superpotential
$W(\phi,\xi)$  does not satisfy a quasi-homogeneous condition for SU(N).
However, one can argue that the supermanifold $\mathcal{M}^{m|n}$ may be 
a kind of super weighted projective space both for the U(1) gauged 
linear sigma model and the U(N) gauged one. 

In the U(1) part, the Calabi-Yau supermanifold must have the same 
number of even coordinates and odd coordinates from Eq. (\ref{q1}).
In the SU(N) part, we must take care to constructing the Calabi-Yau 
supermanifold. Because of the more stringent conditions to be imposed on
the form of the superpotential than in the U(1) case, it seems define 
a certain kind of new supermanifold other than $WCP^{m-1|n}$, which we 
cannot identify exactly among mathematically defined objects.
In our forthcoming paper, we intend to investigate the relationships 
between the non-linear sigma model and (0,2) linear sigma model in order to 
investigate further the correspondence with the super Landau-Ginzburg theory.
Then, we expect to establish the correspondence between the $D=2$, $(0,2)$ 
gauged linear sigma model 
in the U(1) and U(N) gauge groups on the supermanifold to the super 
Landau-Ginzburg model at $r\ll 0$, 
which has been reported in the $D=2$, $(2,2)$ U(1) gauged linear 
sigma model\cite{aganagic1,witten2}.
As a second step, we hope to investigate the Calabi-Yau supermanifold on the 
constructed U(1) and U(N) gauged linear sigma model mathematically 
\cite{Grassi1,Catenacci1}, by using the super Landau-Ginzburg mirror symmetry.



\end{document}